\PassOptionsToPackage{unicode}{hyperref}
\PassOptionsToPackage{hyphens}{url}
\PassOptionsToPackage{dvipsnames,svgnames,x11names}{xcolor}
\documentclass[
]{article}
\usepackage{amsmath,amssymb}
\usepackage{lmodern}
\usepackage{iftex}
\ifPDFTeX
  \usepackage[T1]{fontenc}
  \usepackage[utf8]{inputenc}
  \usepackage{textcomp} 
\else 
  \usepackage{unicode-math}
  \defaultfontfeatures{Scale=MatchLowercase}
  \defaultfontfeatures[\rmfamily]{Ligatures=TeX,Scale=1}
\fi
\IfFileExists{upquote.sty}{\usepackage{upquote}}{}
\IfFileExists{microtype.sty}{
  \usepackage[]{microtype}
  \UseMicrotypeSet[protrusion]{basicmath} 
}{}
\makeatletter
\@ifundefined{KOMAClassName}{
  \IfFileExists{parskip.sty}{%
    \usepackage{parskip}
  }{
    \setlength{\parindent}{0pt}
    \setlength{\parskip}{6pt plus 2pt minus 1pt}}
}{
  \KOMAoptions{parskip=half}}
\makeatother
\usepackage{xcolor}
\usepackage{graphicx}
\makeatletter
\def\maxwidth{\ifdim\Gin@nat@width>\linewidth\linewidth\else\Gin@nat@width\fi}
\def\maxheight{\ifdim\Gin@nat@height>\textheight\textheight\else\Gin@nat@height\fi}
\makeatother
\setkeys{Gin}{width=\maxwidth,height=\maxheight,keepaspectratio}
\makeatletter
\def\fps@figure{htbp}
\makeatother
\usepackage{svg}
\providecommand{\tightlist}{%
  \setlength{\itemsep}{0pt}\setlength{\parskip}{0pt}}
\NewDocumentCommand\citeproctext{}{}
\NewDocumentCommand\citeproc{mm}{%
  \begingroup\def\citeproctext{#2}\cite{#1}\endgroup}
\makeatletter
 \let\@cite@ofmt\@firstofone
 \def\@biblabel#1{}
 \def\@cite#1#2{{#1\if@tempswa , #2\fi}}
\makeatother
\newlength{\cslhangindent}
\newlength{\csllabelwidth}
\newenvironment{CSLReferences}[2] 
 {\begin{list}{}{%
  \setlength{\itemindent}{0pt}
  \setlength{\leftmargin}{0pt}
  \setlength{\parsep}{0pt}
  \ifodd #1
   \setlength{\leftmargin}{\cslhangindent}
   \setlength{\itemindent}{-1\cslhangindent}
  \fi
  \setlength{\itemsep}{#2\baselineskip}}}
 {\end{list}}
\usepackage{calc}

\ifLuaTeX
\usepackage[bidi=basic]{babel}
\else
\usepackage[bidi=default]{babel}
\fi
\babelprovide[main,import]{american}

\def\languageshorthands#1{}
\ifLuaTeX
  \usepackage{selnolig}  
\fi
\IfFileExists{bookmark.sty}{\usepackage{bookmark}}{\usepackage{hyperref}}
\IfFileExists{xurl.sty}{\usepackage{xurl}}{} 
\hypersetup{
  pdftitle={NPAP: Network Partitioning and Aggregation Package for
Python},
  pdfauthor={Marco Anarmo, Benjamin Stöckl, Yannick Werner, Sonja
Wogrin},
  pdflang={en-US},
  colorlinks=true,
  linkcolor={Maroon},
  filecolor={Maroon},
  citecolor={Blue},
  urlcolor={Blue},
  pdfcreator={LaTeX via pandoc}}

\title{NPAP: Network Partitioning and Aggregation Package for Python}

\definecolor{c53baa1}{RGB}{83,186,161}
\definecolor{c202826}{RGB}{32,40,38}

\usepackage[affil-it]{authblk}
\usepackage{orcidlink}
\author[1,2%
  ]{Marco Anarmo%
    \,\orcidlink{0009-0000-3806-7946}\,%
    }
\author[1,2%
  ]{Benjamin Stöckl%
    \,\orcidlink{0009-0005-6579-8169}\,%
    }
\author[1,2%
  ]{Yannick Werner%
    \,\orcidlink{0000-0002-6674-805X}\,%
    }
\author[1,2%
  ]{Sonja Wogrin%
    \,\orcidlink{0000-0002-3889-7197}\,%
    }

\affil[1]{Institute of Electricity Economics and Energy Innovation
(IEE), Graz University of Technology, Inffeldgasse 18, Graz, Austria%
  }
\affil[2]{Research Center ENERGETIC, Graz University of Technology,
Rechbauerstraße 12, Graz, Austria%
  }
\date{10 April 2026}

\begin{document}
\maketitle

\section{Summary}\label{summary}

NPAP (Network Partitioning and Aggregation Package) is an open-source
Python library for reducing the spatial complexity of network graphs.
Built on NetworkX (\citeproc{ref-NetworkX}{Hagberg et al., 2008}), it
provides an accessible standalone package designed to be readily
integrated with other software and frameworks. Instead of treating the
spatial reduction process as a single action, NPAP explicitly splits it
into two distinct steps: partitioning, which assigns vertices (nodes) to
groups (clusters), and aggregation, which reduces the network based on a
given assignment. NPAP's strategy pattern architecture allows users to
employ and register custom partitioning and aggregation strategies
seamlessly without modifying the core code. Currently, NPAP provides 13
different partitioning strategies and two pre-defined aggregation
profiles. Although initially developed with a focus on power systems,
its architecture is general-purpose and applicable to any network graph.

\section{Statement of need}\label{statement-of-need}

Real-world electricity grids have grown significantly in size and
complexity, potentially spanning many thousands of nodes and edges.
Consequently, energy system optimization models representing those
networks have become computationally intractable and challenging to
solve. To regain tractability and computational efficiency, modelers
frequently apply temporal and spatial aggregation techniques
(\citeproc{ref-Kotzur2021}{Kotzur et al., 2021}). Today, temporal
complexity reduction, known as time series aggregation, is a
well-established field (\citeproc{ref-Hoffmann2020}{Hoffmann et al.,
2020}; \citeproc{ref-Teichgraeber2022}{Teichgraeber \& Brandt, 2022};
\citeproc{ref-Wogrin2023}{Wogrin, 2023}). Tools like tsam (Time Series
Aggregation Module) (\citeproc{ref-Hoffmann2022}{Hoffmann et al., 2022})
have consolidated multiple temporal aggregation algorithms into a
single, reusable Python library that has become a standard component in
many energy system optimization modeling frameworks. Before tsam,
modelers implemented these methods individually and ad hoc for each
project, challenging reusability and comparability.

This is exactly where the research on spatial complexity reduction is
at: While there is some research on network partitioning
(\citeproc{ref-frysztacki_comparison_2022}{Frysztacki et al., 2022}) and
aggregation (\citeproc{ref-colonetti_ward_2025}{Colonetti, 2025})
methods available, there exists no standalone tool that brings all
methods together and is easy to use, extend, and does not rely on
framework-specific data structures. NPAP fills this gap as a standalone
package that works with any NetworkX (\citeproc{ref-NetworkX}{Hagberg et
al., 2008}) graph. The target audience includes power systems
researchers, energy modelers, network analysts, and more broadly, anyone
interested in reducing the spatial complexity of a graph.

\section{State of the field}\label{state-of-the-field}

Within the energy system optimization community, existing
implementations of network partitioning and aggregation methods are
usually tightly rooted to the respective frameworks' internal data
structure and cannot be reused independently or by other tools and
frameworks. The built-in spatial clustering module of PyPSA
(\citeproc{ref-Brown2018a}{Brown et al., 2018}), for example, provides
busmap-based spatial aggregation using methods such as k-means or
hierarchical clustering on geographical coordinates of buses, based on
PyPSA's \texttt{Network} object and \texttt{buses} DataFrame. As another
example, ETHOS.FINE (\citeproc{ref-Kluetz2025}{Klütz et al., 2025})
implements spatial reduction through string-, distance-, and
parameter-based partitioning modes (\citeproc{ref-Patil2022}{Patil et
al., 2022}), which are strictly tied to the framework's internal
region-based data model. Researchers working with other frameworks or
custom network representations cannot leverage these implementations
without significant adaptation to the core code structure.

The analogy with temporal complexity reduction and the tsam package
(\citeproc{ref-Hoffmann2022}{Hoffmann et al., 2022}) is instructive.
NPAP extends this paradigm to the spatial dimension. Its unique
contributions include:

\begin{itemize}
\tightlist
\item
  Standalone Architecture: A pip-installable library decoupled from any
  specific energy system modeling framework.
\item
  Extensibility: A strategy pattern architecture that allows new
  functionalities to be added without modifying core code.
\item
  Pipeline Control: An explicit conceptual separation between
  partitioning and aggregation, enabling rigorous control over each
  stage of the reduction pipeline.
\item
  Advanced Physical Constraints: Built-in support for voltage-aware and
  AC-island-aware partitioning for real-world electricity networks.
\end{itemize}

\section{Pipeline architecture}\label{pipeline-architecture}

\begin{figure}
\centering
\includegraphics[width=\linewidth,height=1\textheight,keepaspectratio]{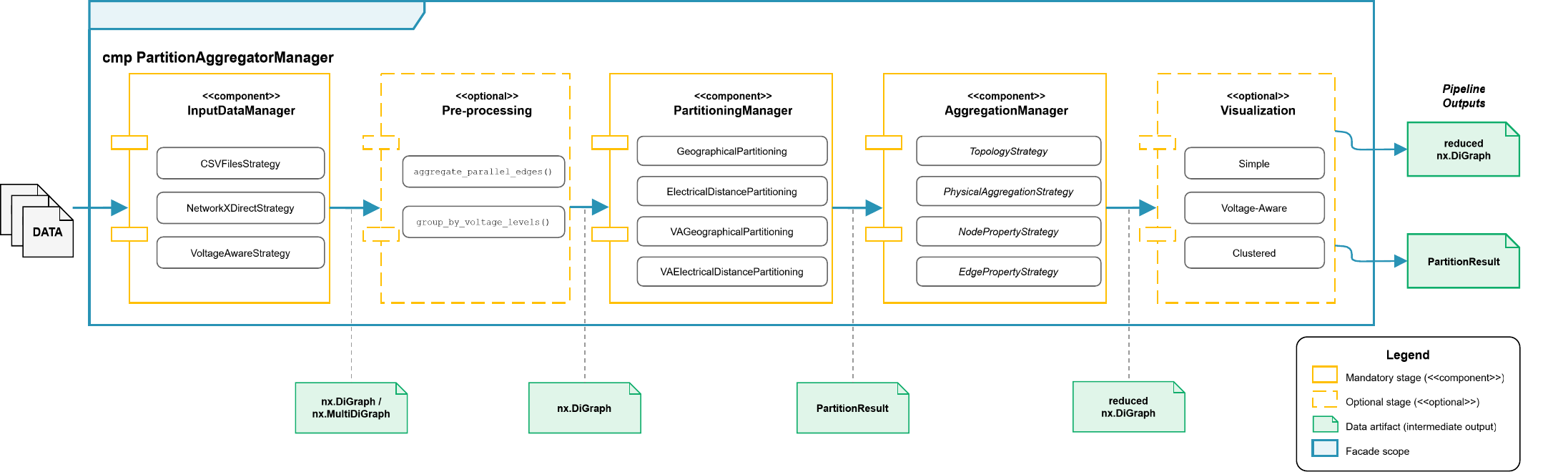}
\caption{NPAP pipeline architecture. \label{fig:pipeline}}
\end{figure}

The full NPAP pipeline is shown in Figure \ref{fig:pipeline}. Initially,
NPAP performs two stages, preparing the network graph. In the first
stage, data is loaded and a NetworkX graph --- used for the reduction
process --- is created and validated. In the second stage, NPAP provides
optional pre-processing steps that prepare the network graph for the
reduction process. The network can be illustrated throughout the
pipeline using the visualization component.

A key design decision in NPAP is the explicit separation of the network
reduction process into two stages:

\begin{enumerate}
\def\labelenumi{\arabic{enumi}.}
\tightlist
\item
  Partitioning: A partitioning strategy maps each node in the original
  network to a node in the aggregated network. This step only determines
  group ``membership''; it does not modify the graph itself.
\item
  Aggregation: An aggregation strategy reduces the network topology
  based on a given partitioning result (mapping) by aggregating nodes,
  edges, and their associated properties.
\end{enumerate}

This separation was identified early in the design process as a
fundamental distinction. It gives users fine-grained control: they can
swap partitioning algorithms independently of the aggregation method,
apply different aggregation strategies to the same partitioning result,
compose custom pipelines that combine strategies from different domains,
or simply use any of them in isolation. The next section explains the
software design in more detail.

\section{Software design}\label{software-design}

\begin{figure}
\centering
\includegraphics[width=\linewidth,height=1\textheight,keepaspectratio]{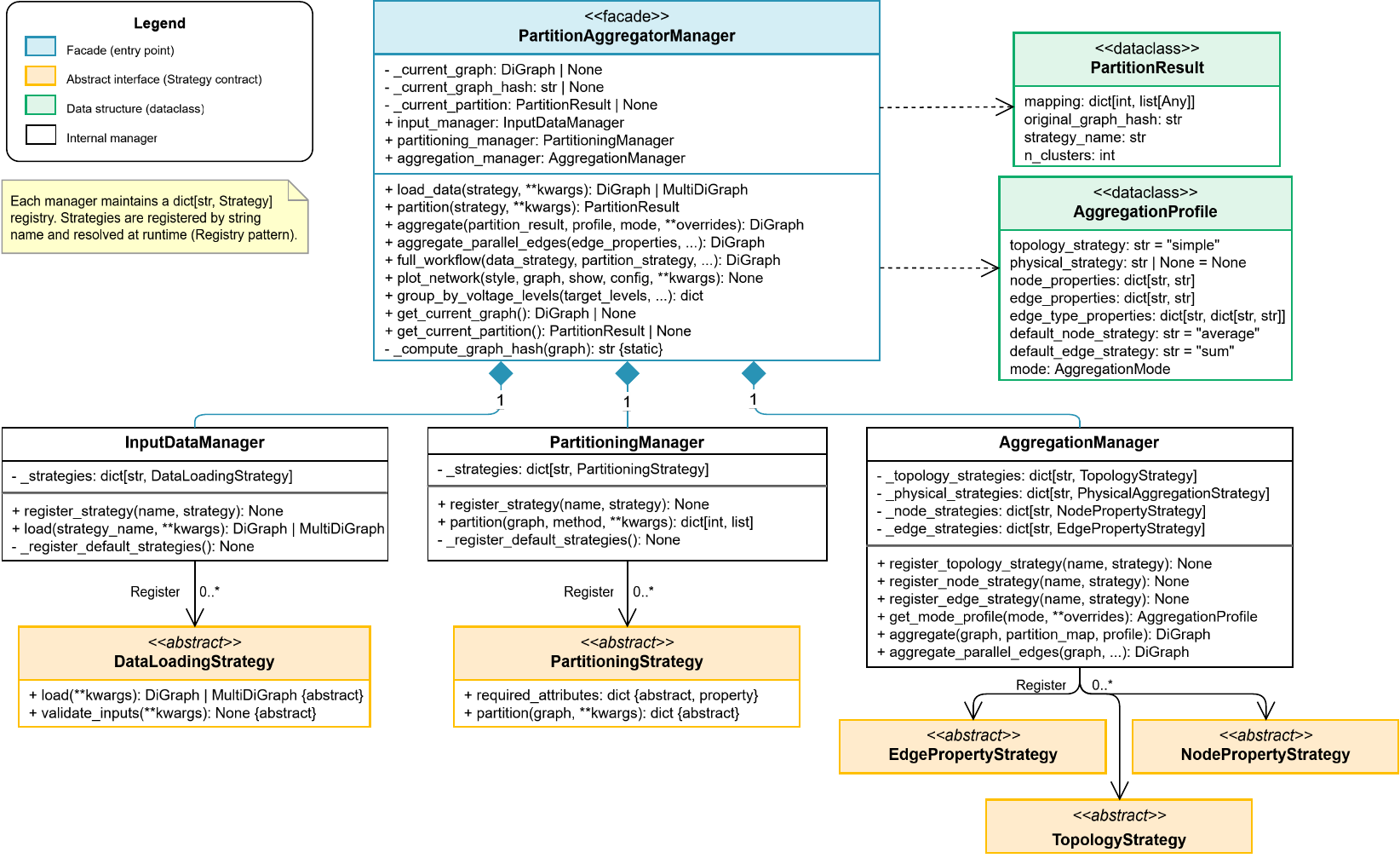}
\caption{NPAP general class diagram. \label{fig:design}}
\end{figure}

A broad overview of NPAP's design is shown in Figure \ref{fig:design}.
The whole workflow is orchestrated and accessed by the
\texttt{PartitionAggregatorManager}. NPAP follows a strategy pattern
with four categories --- data loading, partitioning, topology
aggregation, and property aggregation --- all orchestrated by three
manager classes, shown at the bottom of the Figure. New strategies
inherit from abstract base classes and register with their respective
managers, enabling users to seamlessly add custom strategies without
modifying core code. To reduce entry barriers for contributing to NPAP,
the whole library is built on well-established Python packages, such as
NetworkX (\citeproc{ref-NetworkX}{Hagberg et al., 2008}) for graph
representation, NumPy (\citeproc{ref-Harris2020}{Harris et al., 2020}),
SciPy (\citeproc{ref-Virtanen2020}{Virtanen et al., 2020}), and
scikit-learn (\citeproc{ref-scikit-learn}{Pedregosa et al., 2011}) for
numerical computation, and Plotly (\citeproc{ref-plotly}{Inc., 2015})
for interactive map-based visualization of results. In the following
sections, we introduce the loading and pre-processing of input data,
partitioning, and the aggregation processes in more detail.

\subsection{Data loading and
pre-processing}\label{data-loading-and-pre-processing}

As NPAP is designed to work with NetworkX graphs, they can be directly
passed as input data, facilitating an easy integration into other
frameworks or software. Besides, NPAP currently supports importing data
from CSV files with two main strategies. The first one works with
general node and edge data targeting generic graph structures. The
second one is a domain-dependent strategy focused on power grids and
includes buses, lines, transformers, converters, and DC links.
User-specific data loading strategies can be registered straightforward
as shown in Figure \ref{fig:design}. Afterward, optional pre-processing
steps are carried out through the \texttt{PartitionAggregatorManager},
such as the aggregation of parallel edges. For power grids,
voltage-level grouping separates the network into independent sub-graphs
per voltage level, enabling voltage-aware partitioning strategies.

\subsection{Partitioning strategies}\label{partitioning-strategies}

NPAP currently provides four families of partitioning strategies
combining geographical and electrical node distance with the option of
partitioning voltage levels independently (voltage-awareness). Each
family supports 13 different partitioning strategies, such as k-means
(\citeproc{ref-lloyd1982least}{Lloyd, 1982}), k-medoids
(\citeproc{ref-schubert2021fast}{Schubert \& Rousseeuw, 2021}), DBSCAN
(\citeproc{ref-ester1996density}{Ester et al., 1996}), HDBSCAN
(\citeproc{ref-campello2013density}{Campello et al., 2013}), and
hierarchical clustering (\citeproc{ref-ward1963hierarchical}{Ward Jr,
1963}). Geographical distance captures latitudes and longitudes and is
suitable for any geo-referenced network. Electrical distance computes
Power Transfer Distribution Factors (PTDFs)
(\citeproc{ref-wood2012power}{Wood et al., 2012}) to capture the
electrical behavior of a power grid rather than its geographical
topology.

In power grids, NPAP automatically detects alternating current (AC)
islands linked solely by DC interconnections and partitions them
independently. This, and the voltage-aware partitioning, is achieved by
setting the distance matrix entries of nodes in different AC islands and
voltage levels, respectively, to infinity. Both approaches are
algorithm-agnostic, i.e, they work with any distance-based partitioning
method without requiring modifications to the algorithm itself. The
partitioning outcome, along with other metadata, is then stored in the
PartitionResult.

\subsection{Three-tier aggregation strategy
pipeline}\label{three-tier-aggregation-strategy-pipeline}

The aggregation process is decomposed into three sequential steps. The
first (topology) tier builds a new network graph based on the mapping
result of the partitioning process by creating a representative node for
each cluster and adjusting the edges between them accordingly.
Afterward, to cover use cases we do not foresee and usability outside
the power system domain, we have included an optional second
(domain-specific) tier. There, users can modify physical properties of
the network graph, e.g., adding new edges that did not exist in the
original graph or explicitly modifying certain properties, such as line
reactances. In the third (property aggregation) tier, user-specified
functions, e.g, sum, average, or equivalent reactances, are then applied
to the remaining node and edge properties. These property aggregation
strategies can be configured by the user in detail through the
aggregation profiles shown in Figure \ref{fig:design}. At the moment,
NPAP provides two pre-defined aggregation profiles.

\section{\texorpdfstring{Research impact statement
\label{sec:impact}}{Research impact statement }}\label{research-impact-statement}

NPAP is designed as a standalone, pip-installable open-source Python
package, released under the MIT license. The library is fully documented
on ReadTheDocs and includes an automated test suite with continuous
integration across Python 3.10, 3.11, and 3.12. It has been tested on
small- and large-scale networks with up to a few thousand nodes.

One potential application of NPAP is within energy system modeling
frameworks. For illustration purposes, we have integrated it into the
well-established PyPSA-Eur (\citeproc{ref-Hoersch2018}{Hörsch et al.,
2018}) framework through an open pull request introducing NPAP as an
alternative to PyPSA's native spatial clustering backend.\footnote{https://github.com/PyPSA/PyPSA/pull/1568}
Thereby showcasing NPAP's framework-agnostic design and highlighting the
minimal adoption effort.

We further utilized this implementation to demonstrate NPAP's usability
and scalability by applying it to the full pan-European transmission
network (\citeproc{ref-Xiong2025}{Xiong et al., 2025}), comprising
around 6800 nodes and 17500 edges. Leveraging the voltage-aware
partitioning strategy with geographical distance and k-means clustering
(\citeproc{ref-lloyd1982least}{Lloyd, 1982}), we analyze the required
infrastructure investments for power lines on different voltage levels
and transformers using PyPSA (\citeproc{ref-Brown2018a}{Brown et al.,
2018}). The work has been submitted to the International Conference on
the European Energy Market 2026.

Active development directions include the extension of the available
partitioning strategies, e.g., the Adjacent Node Agglomerative
Clustering algorithm (\citeproc{ref-Stoeckl2025a}{Stöckl et al., 2025}),
and aggregation strategies, e.g., based on PTDFs
(\citeproc{ref-Fortenbacher2018}{Fortenbacher et al., 2018}), as well as
the extension to other physical infrastructures, such as hydrogen
networks, which require distinct strategies. Future extensions include,
among others, a filtering module that allows selecting subsets of the
network, e.g., a country or region, as an additional pre-processing step
before partitioning.

\section{AI Usage Disclosure}\label{ai-usage-disclosure}

Several artificial intelligence (AI) tools were used during the
development of NPAP. Google Gemini (Pro) was used for non-code-related
tasks, such as learning about power systems. Anthropic Claude (Sonnet
4.5, Opus 4.5, and 4.6), accessed through both the desktop application
and the Claude Code terminal interface, was used for code-related
activities, including software design brainstorming, implementation
support, test coverage, or writing the documentation. Grammarly was used
for grammar and spell checking. No other code-generation tools (such as
GitHub Copilot, Codex, or Cursor) were used. All AI-generated output was
carefully reviewed, tested, and frequently modified before inclusion.
The authors take full responsibility for the code.

\section{Acknowledgements}\label{acknowledgements}

The work done on the package by the contributors of the Institute of
Electricity Economics and Energy Innovation (IEE) at Graz University of
Technology was funded by the European Union (ERC, NetZero-Opt,
101116212). Views and opinions expressed are, however, those of the
author(s) only and do not necessarily reflect those of the European
Union or the European Research Council. Neither the European Union nor
the granting authority can be held responsible for them.

\section*{References}\label{references}
\addcontentsline{toc}{section}{References}

\protect\phantomsection\label{refs}


\begin{CSLReferences}{1}{0}
\bibitem[\citeproctext]{ref-Brown2018a}
Brown, T., Hörsch, J., \& Schlachtberger, D. (2018). PyPSA: Python for
power system analysis. \emph{Journal of Open Research Software},
\emph{6}(1), 4. \url{https://doi.org/10.5334/jors.188}

\bibitem[\citeproctext]{ref-campello2013density}
Campello, R. J. G. B., Moulavi, D., \& Sander, J. (2013). Density-based
clustering based on hierarchical density estimates. \emph{Advances in
Knowledge Discovery and Data Mining}, 160--172.
\url{https://doi.org/10.1007/978-3-642-37456-2_14}

\bibitem[\citeproctext]{ref-colonetti_ward_2025}
Colonetti, B. (2025). Ward reduction in unit-commitment problems.
\emph{Electric Power Systems Research}, \emph{240}, 111271.
\url{https://doi.org/10.1016/j.epsr.2024.111271}

\bibitem[\citeproctext]{ref-ester1996density}
Ester, M., Kriegel, H.-P., Sander, J., \& Xu, X. (1996). A density-based
algorithm for discovering clusters in large spatial databases with
noise. \emph{Proceedings of the Second International Conference on
Knowledge Discovery and Data Mining (KDD-96)}, 226--231.

\bibitem[\citeproctext]{ref-Fortenbacher2018}
Fortenbacher, P., Demiray, T., \& Schaffner, C. (2018). Transmission
network reduction method using nonlinear optimization. \emph{2018 Power
Systems Computation Conference (PSCC)}, 1--7.
\url{https://doi.org/10.23919/pscc.2018.8442974}

\bibitem[\citeproctext]{ref-frysztacki_comparison_2022}
Frysztacki, M. M., Recht, G., \& Brown, T. (2022). A comparison of
clustering methods for the spatial reduction of renewable electricity
optimisation models of europe. \emph{Energy Inform}, \emph{5}(1), 4.
\url{https://doi.org/10.1186/s42162-022-00187-7}

\bibitem[\citeproctext]{ref-NetworkX}
Hagberg, A. A., Schult, D. A., \& Swart, P. J. (2008). Exploring network
structure, dynamics, and function using NetworkX. In G. Varoquaux, T.
Vaught, \& J. Millman (Eds.), \emph{Proceedings of the 7th python in
science conference} (pp. 11--15).

\bibitem[\citeproctext]{ref-Harris2020}
Harris, C. R., Millman, K. J., Walt, S. J. van der, Gommers, R.,
Virtanen, P., Cournapeau, D., Wieser, E., Taylor, J., Berg, S., Smith,
N. J., Kern, R., Picus, M., Hoyer, S., Kerkwijk, M. H. van, Brett, M.,
Haldane, A., Río, J. F. del, Wiebe, M., Peterson, P., \ldots{} Oliphant,
T. E. (2020). Array programming with NumPy. \emph{Nature},
\emph{585}(7825), 357--362.
\url{https://doi.org/10.1038/s41586-020-2649-2}

\bibitem[\citeproctext]{ref-Hoffmann2022}
Hoffmann, M., Kotzur, L., \& Stolten, D. (2022). The pareto-optimal
temporal aggregation of energy system models. \emph{Applied Energy},
\emph{315}, 119029. \url{https://doi.org/10.1016/j.apenergy.2022.119029}

\bibitem[\citeproctext]{ref-Hoffmann2020}
Hoffmann, M., Kotzur, L., Stolten, D., \& Robinius, M. (2020). A review
on time series aggregation methods for energy system models.
\emph{Energies}, \emph{13}(3), 641.
\url{https://doi.org/10.3390/en13030641}

\bibitem[\citeproctext]{ref-Hoersch2018}
Hörsch, J., Hofmann, F., Schlachtberger, D., \& Brown, T. (2018).
{PyPSA}-{Eur}: An open optimisation model of the european transmission
system. \emph{Energy Strategy Reviews}, \emph{22}, 207--215.
\url{https://doi.org/10.1016/j.esr.2018.08.012}

\bibitem[\citeproctext]{ref-plotly}
Inc., P. T. (2015). \emph{Collaborative data science}. Plotly
Technologies Inc. \url{https://plot.ly}

\bibitem[\citeproctext]{ref-Kluetz2025}
Klütz, T., Knosala, K., Behrens, J., Maier, R., Hoffmann, M., Pflugradt,
N., \& Stolten, D. (2025). ETHOS.FINE: A framework for integrated energy
system assessment. \emph{Journal of Open Source Software},
\emph{10}(105), 6274. \url{https://doi.org/10.21105/joss.06274}

\bibitem[\citeproctext]{ref-Kotzur2021}
Kotzur, L., Nolting, L., Hoffmann, M., Groß, T., Smolenko, A.,
Priesmann, J., Büsing, H., Beer, R., Kullmann, F., Singh, B.,
Praktiknjo, A., Stolten, D., \& Robinius, M. (2021). A modeler's guide
to handle complexity in energy systems optimization. \emph{Advances in
Applied Energy}, \emph{4}, 100063.
\url{https://doi.org/10.1016/j.adapen.2021.100063}

\bibitem[\citeproctext]{ref-lloyd1982least}
Lloyd, S. P. (1982). Least squares quantization in PCM. \emph{IEEE
Transactions on Information Theory}, \emph{28}(2), 129--137.
\url{https://doi.org/10.1109/TIT.1982.1056489}

\bibitem[\citeproctext]{ref-Patil2022}
Patil, S., Kotzur, L., \& Stolten, D. (2022). Advanced spatial and
technological aggregation scheme for energy system models.
\emph{Energies}, \emph{15}(24), 9517.
\url{https://doi.org/10.3390/en15249517}

\bibitem[\citeproctext]{ref-scikit-learn}
Pedregosa, F., Varoquaux, G., Gramfort, A., Michel, V., Thirion, B.,
Grisel, O., Blondel, M., Prettenhofer, P., Weiss, R., Dubourg, V.,
Vanderplas, J., Passos, A., Cournapeau, D., Brucher, M., Perrot, M., \&
Duchesnay, E. (2011). Scikit-learn: Machine learning in {P}ython.
\emph{Journal of Machine Learning Research}, \emph{12}, 2825--2830.

\bibitem[\citeproctext]{ref-schubert2021fast}
Schubert, E., \& Rousseeuw, P. J. (2021). Fast and eager k-medoids
clustering: O(k) runtime improvement of the PAM, CLARA, and CLARANS
algorithms. \emph{Information Systems}, \emph{101}, 101804.
\url{https://doi.org/10.1016/j.is.2021.101804}

\bibitem[\citeproctext]{ref-Stoeckl2025a}
Stöckl, B., Werner, Y., \& Wogrin, S. (2025). Congestion-sensitive grid
aggregation for DC optimal power flow. \emph{2025 IEEE Kiel PowerTech},
1--7. \url{https://doi.org/10.1109/powertech59965.2025.11180585}

\bibitem[\citeproctext]{ref-Teichgraeber2022}
Teichgraeber, H., \& Brandt, A. R. (2022). Time-series aggregation for
the optimization of energy systems: Goals, challenges, approaches, and
opportunities. \emph{Renewable and Sustainable Energy Reviews},
\emph{157}, 111984. \url{https://doi.org/10.1016/j.rser.2021.111984}

\bibitem[\citeproctext]{ref-Virtanen2020}
Virtanen, P., Gommers, R., Oliphant, T. E., Haberland, M., Reddy, T.,
Cournapeau, D., Burovski, E., Peterson, P., Weckesser, W., Bright, J.,
Walt, S. J. van der, Brett, M., Wilson, J., Millman, K. J., Mayorov, N.,
Nelson, A. R. J., Jones, E., Kern, R., Larson, E., \ldots{}
Vázquez-Baeza, Y. (2020). SciPy 1.0: Fundamental algorithms for
scientific computing in python. \emph{Nature Methods}, \emph{17}(3),
261--272. \url{https://doi.org/10.1038/s41592-019-0686-2}

\bibitem[\citeproctext]{ref-ward1963hierarchical}
Ward Jr, J. H. (1963). Hierarchical grouping to optimize an objective
function. \emph{Journal of the American Statistical Association},
\emph{58}(301), 236--244.
\url{https://doi.org/10.1080/01621459.1963.10500845}

\bibitem[\citeproctext]{ref-Wogrin2023}
Wogrin, S. (2023). Time series aggregation for optimization:
One-size-fits-all? \emph{IEEE Transactions on Smart Grid}, \emph{14}(3),
2489--2492. \url{https://doi.org/10.1109/TSG.2023.3242467}

\bibitem[\citeproctext]{ref-wood2012power}
Wood, A. J., Wollenberg, B. F., \& Sheblé, G. B. (2012). \emph{Power
generation, operation, and control} (3rd ed.). John Wiley \& Sons.

\bibitem[\citeproctext]{ref-Xiong2025}
Xiong, B., Fioriti, D., Neumann, F., Riepin, I., \& Brown, T. (2025).
Modelling the high-voltage grid using open data for europe and beyond.
\emph{Scientific Data}, \emph{12}, 277.
\url{https://doi.org/10.1038/s41597-025-04550-7}

\end{CSLReferences}
\end{document}